\documentclass{elsart1p}

\newcommand{\be}{\begin{equation}}
\newcommand{\ee}{\end{equation}}
\newcommand{\ba}{\begin{eqnarray}}
\newcommand{\ea}{\end{eqnarray}}
\newcommand{\ban}{\begin{eqnarray*}}
\newcommand{\ean}{\end{eqnarray*}}

\usepackage{amssymb}

\begin{document}

\begin{frontmatter}

\title{Fluctuations of Chromodynamic Fields \\ in Quark-Gluon 
Plasma\thanksref{label1}}

\thanks[label1]{Talk given at the {\it Conference on Strong and 
Electroweak Matter 2008} (SEWM08), August 26-29, 2008, 
Amsterdam, The Netherlands}

\author{Stanis\l aw Mr\' owczy\' nski}

\address{Institute of Physics, Jan Kochanowski University, 
ul.~\'Swi\c etokrzyska 15, 25-406 Kielce, Poland \\
and So\l tan Institute for Nuclear Studies, 
ul.~Ho\.za 69, 00-681 Warsaw, Poland}

\begin{abstract}

Chromodynamic fluctuations in the collisionless quark-gluon plasma 
are found as a solution of the initial value linearized problem.
The stable and unstable plasmas are discussed. 

\end{abstract}

\end{frontmatter}


In the quark-gluon plasma (QGP), which is on average locally colorless,
chromodynamic fields, color charges and currents experience random 
fluctuations which appear to influence dynamics of the whole system. 
In the equilibrium plasmas there are characteristic stationary spectra 
of fluctuations which can be found by means of the fluctuation-dissipation
relations. Fluctuation spectra in nonequilibrium systems evolve in time, 
and their characteristics usually depend on an initial state of the 
system. Our aim is to discuss chromodynamic fluctuations in equilibrium 
and nonequilibrium QGP. We are particularly interested in QGP produced 
at the early stage of relativistic heavy-ion collisions. Such a plasma 
is unstable with respect to chromomagnetic modes due to anisotropic 
momentum distribution of quarks and gluons (partons), see the review 
\cite{Mrowczynski:2005ki}. The instability growth is associated with 
generation of chromomagnetic fields which in turn influence various 
plasma properties. In particular, transport coefficients of such a 
plasma, which are controlled by the fluctuation spectrum of 
chromomagnetic fields, are then strongly modified \cite{Asakawa:2006jn}. 

Fluctuations can be theoretically studied by means of several methods 
reviewed in the classical monographs \cite{Akh75,Sit82}. We choose
the method which is clearly exposed in the handbook \cite{LP81}. 
The method - applicable to both equilibrium and nonequilibrium plasmas 
- provides the spectrum of fluctuations as a solution of the initial 
value (linearized) problem. The initial plasma state is assumed to be 
on average charge neutral, stationary and homogeneous. When the state 
is stable, the initial fluctuations are explicitly shown to exponentially 
decay and in the long time limit one finds a stationary spectrum of the 
fluctuations. When the initial state is unstable, the memory of initial 
fluctuations is not lost, as the unstable modes, which are usually present 
in the initial fluctuation spectrum, exponentially grow. The fluctuations 
of chromodynamic fields in QGP have been discussed in detail in our study 
\cite{Mrowczynski:2008ae}. Here we briefly summarize the results.

The transport theory of weakly coupled quark-gluon plasma, which 
forms the basis  of our analysis, is formulated in terms of particles and 
classical fields. The particles - quarks, antiquarks and gluons - 
should be understood as sufficiently hard quasiparticle excitations 
of QCD quantum fields while the classical fields are highly 
populated soft gluonic modes. The transport equation of quarks reads
\be
\label{transport-eq}
\big(D^0 + {\bf v} \cdot {\bf D} \big) Q(t,{\bf r},{\bf p})
- {g \over 2}
\{{\bf E}(t,{\bf r}) + {\bf v} \times {\bf B}(t,{\bf r}), 
\nabla_p Q(t,{\bf r},{\bf p}) \}
= 0 \;, 
\ee
where $Q(t,{\bf r},{\bf p})$ is the on-mss-shell quark distribution 
function which is $N_c\times N_c$ hermitean matrices belonging to 
the fundamental representation of the SU($N_c$) group; the covariant 
derivative in the four-vector notation reads 
$D^\mu \equiv \partial^\mu - ig [A^\mu(x),\cdots \;]$ and 
${\bf E}(t,{\bf r})$ and ${\bf B}(t,{\bf r})$ are the chromoelectric 
and chromomagnetic fields. The symbol $\{\dots , \dots \}$ denotes 
the anticommutator. Since the fluctuations of interest are assumed 
to be of the time scale, which is much shorter than that of inter-parton 
collisions, the collision terms are absent in Eq.~(\ref{transport-eq}). 
There are analogous transport equations for antiquark 
($\bar Q(t,{\bf r},{\bf p})$) and gluon ($G(t,{\bf r},{\bf p})$) 
distribution functions.

The transport equations are supplemented by the Yang-Mills equations 
describing a self-consistent generation of the chromoelectric and 
chromomagnetic fields by the color four-current $j^\mu =(\rho , {\bf j})$ 
$$
j^\mu_a (t,{\bf r}) = - g \int {d^3 p \over (2\pi)^3} \,
\frac{p^\mu}{E_{\bf p}}
{\rm Tr}\Big[\tau^a \big(Q (t,{\bf r},{\bf p})
- \bar Q(t,{\bf r},{\bf p}) \big)
+ T^a G(t,{\bf r},{\bf p}) \Big] \;,
$$
where $\tau^a$, $T^a$ with $a = 1, ... \, ,N_c^2-1$ are the SU($N_c$)
group generators in the fundamental and adjoint representations. 

We consider small deviations from a stationary homogeneous 
state which is globally and locally colorless; there are no currents 
as well. The quark distribution function of this state is 
$Q^0_{nm}({\bf p}) = n ({\bf p}) \: \delta^{nm}$. Due to the 
absence of color charges and currents in the stationary and 
homogeneous state, the chromoelectric ${\bf E}(t,{\bf r})$ and 
chromomagnetic ${\bf B}(t,{\bf r})$ fields are expected to vanish
while the potentials $A^0(t,{\bf r}), {\bf A}(t,{\bf r})$ are of 
pure gauge only. Since the plasma under considerations is assumed
to be weakly coupled with the perturbative vacuum state, the 
potentials can be gauge away to vanish. 

We write down the quark distribution function as
$Q(t,{\bf r},{\bf p}) =  Q^0({\bf p}) + 
\delta Q(t,{\bf r},{\bf p})$, and we assume that
$|Q^0| \gg |\delta Q| $ and $|\nabla_p Q^0| \gg | \nabla_p \delta Q|$
with the analogous formulas for antiquarks and gluons.
We linearize the transport (\ref{transport-eq}) and Yang-Mills
equations in the deviations from the stationary homogeneous 
state. We assume that $\delta Q$, ${\bf E}$, ${\bf B}$, $A^0$ 
and ${\bf A}$ are all of the same order. The linearized transport 
equation is
$$
\Big(\frac{\partial}{\partial t} + {\bf v} \cdot \nabla \Big) 
\delta Q(t,{\bf r},{\bf p})
- g \big({\bf E}(t,{\bf r}) + {\bf v} \times {\bf B}(t,{\bf r})\big) 
\nabla_p n({\bf p})
= 0 \;. 
$$
The Yang-Mills equations get after the linearization the familiar 
form of Maxwell equations of multi-component electrodynamics. 

The linearized transport and Maxwell equations are solved
with the initial conditions 
$\delta Q(t=0,{\bf r},{\bf p}) = \delta Q_0({\bf r},{\bf p})$,
${\bf E}(t=0,{\bf r}) = {\bf E}_0({\bf r})$,
${\bf B}(t=0,{\bf r})= {\bf B}_0({\bf r})$,
by means of the one-sided Fourier transformation defined as
$$
f(\omega,{\bf k}) = \int_0^\infty dt \int d^3r 
e^{i(\omega t - {\bf k}\cdot {\bf r})}
f(t,{\bf r}) \;.
$$
The chromoelectric field, which solves the equations, is found as
\ba
\nonumber
\big[ - {\bf k}^2 \delta^{ij} &+& k^ik^j 
+ \omega^2 \varepsilon^{ij}(\omega,{\bf k}) \big] E^j_a(\omega,{\bf k})
= - i\frac{g^2}{2} \int {d^3p \over (2\pi)^3} \,
\frac{v^i \big({\bf v}\times {\bf B}_{a0}({\bf k})\big)^j
\nabla_p^j f({\bf p})}{\omega - {\bf v}\cdot {\bf k}}
\\[2mm]
\label{E-field2} 
&-& g \omega \int {d^3p \over (2\pi)^3} \,
\frac{v^i} {\omega - {\bf k}\cdot {\bf v}}\, 
\delta N^a_0({\bf k},{\bf p})
+ i \omega E_{a0}^i({\bf k})
-i \big({\bf k} \times {\bf B}_{a0}({\bf k})\big)^i \;,
\ea
where $f({\bf p}) \equiv n({\bf p}) + \bar n ({\bf p})
+ 2N_c n_g({\bf p})$ and $\delta N^a_0({\bf r},{\bf p}) \equiv
{\rm Tr}\big[\tau^a \big(\delta Q_0 ({\bf r},{\bf p})
- \delta \bar Q_0({\bf r},{\bf p}) \big)
+ T^a \delta G_0({\bf r},{\bf p}) \big]$;
$\varepsilon^{ij}(\omega,{\bf k})$ is the chromodielectric 
tensor of, in general, anisotropic plasma in the collisionless limit;
$\varepsilon^{ij}(\omega,{\bf k})$ does not carry any color indices
as it corresponds to a colorless state of the plasma.

When the plasma stationary state is isotropic, the dielectric tensor 
can be expressed through its longitudinal ($\varepsilon_L(\omega,{\bf k})$)
and transverse ($\varepsilon_L(\omega,{\bf k})$) components and the matrix $\Sigma^{ij}(\omega,{\bf k}) 
\equiv - {\bf k}^2 \delta^{ij} + k^ik^j
+ \omega^2 \varepsilon^{ij}(\omega,{\bf k})$ 
from the left-hand-side of Eq.~(\ref{E-field2}) 
can be inverted as 
$$
 (\Sigma^{-1})^{ij}(\omega,{\bf k}) = 
\frac{1}{\omega^2 \varepsilon_L(\omega,{\bf k})}
\frac{k^ik^j}{{\bf k}^2}
+ \frac{1}{\omega^2 \varepsilon_T(\omega,{\bf k})-{\bf k}^2}
\Big(\delta^{ij} - \frac{k^ik^j}{{\bf k}^2}\Big) \;.
$$
Then, Eq.~(\ref{E-field2}) provides an explicit expression of
the chromoelectric field. Using the Maxwell equations, the chromomagnetic 
field, color current and color density can be all expressed through 
the chromoelectric field. 

The correlation functions 
$\langle E^i_a(t_1,{\bf r}_1) E^j_b(t_2,{\bf r}_2) \rangle$,
$\langle B^i_a(t_1,{\bf r}_1) B^j_b(t_2,{\bf r}_2) \rangle$,
where $\langle \cdots \rangle$ denotes averaging over statistical
ensemble, are determined by the initial correlations such as
$\langle \delta N_0^a({\bf r}_1,{\bf p}_1) 
\delta N_0^b({\bf r}_2,{\bf p}_2)\rangle$, 
$\langle E_{a0}^i({\bf r}_1) E_{b0}^j({\bf r}_2) \rangle$,
$\langle \delta N_0^a({\bf r}_1,{\bf p}_1) 
E_{b0}^j({\bf r}_2) \rangle$ which can be all expressed, 
using the Maxwell equations, through the correlation function 
of the distribution functions. The latter one is identified with
the respective correlation function of the classical system of 
free quarks, antiquarks and gluons which on average is stationary 
and homogeneous. For quarks the correlation function is
\ban
\langle \delta Q^{mn}(t_1,{\bf r}_1,{\bf p}_1) 
\delta Q^{pr}(t_2,{\bf r}_2,{\bf p}_2)\rangle_{\rm free}
&=& \delta^{mr} \delta^{np} 
(2\pi )^3 \delta^{(3)}({\bf p}_1 - {\bf p}_2) \,
\\[2mm] \nonumber
&\times&
\delta^{(3)}\big({\bf r}_2 - {\bf r}_1 
- {\bf v}_1(t_2 - t_1)\big) \: n({\bf p}_1) \;.
\ean
 
In the case of equilibrium plasma where all collective modes
are damped, we consider the times which are much longer than 
the decay time of collective excitations. Then, the correlation 
function of the chromoelectric fields equals
$$
\langle E_a^i(t_1,{\bf r}_1) E_b^j(t_2,{\bf r}_2) \rangle_\infty 
= 
\int {d\omega \over 2\pi} {d^3k \over (2\pi)^3}
e^{-i \big(\omega (t_1 - t_2)
 - {\bf k}\cdot ({\bf r}_1 - {\bf r}_2)\big)} 
\langle E_a^i E_b^j\rangle_{\omega \, {\bf k}} \;,
$$ 
where the fluctuation spectrum is
$$
\langle E_a^i E_b^j\rangle_{\omega \, {\bf k}}
= 
2 \delta^{ab} T \omega^3 \bigg[
\frac{k^ik^j}{{\bf k}^2}
\frac{\Im \varepsilon_L(\omega,{\bf k})}
{|\omega^2 \varepsilon_L(\omega,{\bf k})|^2}
+
\Big(\delta^{ij} - \frac{k^ik^j}{{\bf k}^2}\Big)
\frac{\Im \varepsilon_T(\omega,{\bf k})}
{|\omega^2 \varepsilon_T(\omega,{\bf k})-{\bf k}^2|^2}
\bigg] \;.
$$
As seen, the fluctuation spectrum has strong peaks corresponding
to the collective modes.  

As an example of a nonequilibrium system, we discuss 
fluctuations of longitudinal chromoelectric fields in the 
two-stream  system which is unstable with respect to  
longitudinal modes. Nonequlibrium calculations are much more 
difficult than the equilibrium ones. The first problem is to 
invert the matrix $\Sigma^{ij}(\omega,{\bf k})$. In the case 
of longitudinal electric field, which is discussed here, the 
matrix is replaced by the scalar function. The distribution 
function of the two-stream system is chosen to be 
$$
f({\bf p}) = (2\pi )^3 n 
\Big[\delta^{(3)}({\bf p} - {\bf q}) + \delta^{(3)}({\bf p} 
+ {\bf q}) \Big] \;,
$$
where $n$ is the effective parton density in a single stream. 
There are four roots $\pm \omega_{\pm}({\bf k})$ of the dispersion 
equation $\varepsilon_L(\omega,{\bf k}) = 0$. The solution 
$\omega_+({\bf k})$ represents the stable modes and $\omega_-({\bf k})$
corresponds to the well-known two-stream electrostatic instability
for ${\bf k}^2 ({\bf k} \cdot {\bf u})^2 < 2 \mu^2 
\big({\bf k}^2 - ({\bf k} \cdot {\bf u})^2\big)$
where ${\bf u} \equiv {\bf q}/|{\bf q}|$ is the stream velocity
and $\mu^2 \equiv g^2n/(2|{\bf q}|)$. Then, 
$\omega_-({\bf k})=i\gamma_{\bf k}$ with $0 \le \gamma_{\bf k} \in R$.

The correlation function of longitudinal chromoelectric
fields generated by the unstable modes is found as 
\ban
\langle E_a^i(t_1,{\bf r}_1) E_b^i(t_2,{\bf r}_2) 
\rangle_{\rm unstable}
= \frac{g^2}{2}\,\delta^{ab} n 
\int {d^3k \over (2\pi)^3} 
\frac{e^{i {\bf k}({\bf r}_1 - {\bf r}_2)}}{{\bf k}^2}
\frac{\big(\gamma_{\bf k}^2 + ({\bf k} \cdot {\bf u})^2\big)^2}
{(\omega_+^2 - \omega_-^2)^2\gamma_{\bf k}^2}
\\[2mm] 
\times
\Big[
\big(\gamma_{\bf k}^2 + ({\bf k} \cdot {\bf u})^2\big)
\cosh \big(\gamma_{\bf k} (t_1 + t_2)\big)
+
\big(\gamma_{\bf k}^2 - ({\bf k} \cdot {\bf u})^2\big)
\cosh \big(\gamma_{\bf k} (t_1 - t_2)\big) \Big] \;.
\ean
As seen, the correlation function of the unstable system is
invariant with respect to space translations -- it depends on the 
difference $({\bf r}_1 - {\bf r}_2)$ only. The plasma state, which 
is initially on average homogeneous, remains like this in course of
the system's temporal evolution. The time dependence of the correlation 
function is very different from the space dependence. The electric 
fields exponentially grow and so does the correlation function both 
in $(t_1 + t_2)$ and $(t_1 - t_2)$. The fluctuation spectrum also 
evolves in time, as the growth rate of unstable modes is wave-vector 
dependent and after a sufficiently long time the fluctuation spectrum 
is dominated by the fastest growing modes.


\end{document}